\documentclass[prl,twocolumn,superscriptaddress,nobibnotes,letterpaper]{revtex4}

\usepackage{times}
\usepackage[pdftex]{graphicx}
\usepackage[pdftex]{epsfig}
\usepackage{amsmath}
\usepackage{pslatex}
\usepackage{amssymb}
\usepackage{amsfonts}
\usepackage{color}
\usepackage{wrapfig}
\usepackage{eucal}
\usepackage{hhline}
\usepackage{threeparttable}
\usepackage{supertabular}
\usepackage{multirow}
\usepackage{tabularx}

\newcommand{\ki}{\kappa_{\mathrm{i}}}
\newcommand{\ke}{\kappa_{\mathrm{e}}}

\def\be{\begin{equation}}
\def\ee{\end{equation}}
\def\bea{\begin{eqnarray}}
\def\eea{\end{eqnarray}}

\newcolumntype{Y}{>{\centering\arraybackslash}X}

\begin{document}

\pagenumbering{arabic}

\title{Highly efficient coupling from an optical fiber to a nanoscale silicon optomechanical cavity}\thanks{This work was published in Appl.\ Phys.\ Lett. \textbf{103}, 181104 (2013).}

\author{Simon Gr\"oblacher}
\thanks{These authors contributed equally to this work.}
\author{Jeff T. Hill}
\thanks{These authors contributed equally to this work.}
\author{Amir H. Safavi-Naeini}
\thanks{These authors contributed equally to this work.}
\affiliation{Thomas J. Watson, Sr., Laboratory of Applied Physics, California Institute of Technology, Pasadena, CA 91125, USA}
\affiliation{Institute for Quantum Information and Matter, California Institute of Technology, Pasadena, CA 91125, USA}
\author{Jasper Chan}
\affiliation{Thomas J. Watson, Sr., Laboratory of Applied Physics, California Institute of Technology, Pasadena, CA 91125, USA}
\author{Oskar Painter}
\email{opainter@caltech.edu}
\homepage{http://copilot.caltech.edu}
\affiliation{Thomas J. Watson, Sr., Laboratory of Applied Physics, California Institute of Technology, Pasadena, CA 91125, USA}
\affiliation{Institute for Quantum Information and Matter, California Institute of Technology, Pasadena, CA 91125, USA}
\affiliation{Max Planck Institute for the Science of Light, G\"unther-Scharowsky-Stra\ss{}e 1/Bldg. 24, D-91058 Erlangen, Germany}



\begin{abstract}
We demonstrate highly efficient coupling of light from an optical fiber to a silicon photonic crystal optomechanical cavity. The fiber-to-cavity coupling utilizes a compact ($L\approx 25$~$\mu$m) intermediate adiabatic coupler. The optical coupling is lithographically controlled, broadband, relatively insensitive to fiber misalignment, and allows for light to be transferred from an optical fiber to, in principle, any photonic chip with refractive index greater than that of the optical fiber. Here we demonstrate single-sided cavity coupling with a total fiber-to-cavity optical power coupling efficiency of 85\%.
\end{abstract}

\maketitle


Efficient coupling of light from an optical fiber to micro- and nano-scale on-chip optical devices is an important, albeit very challenging, task in integrated photonics~\cite{Bogaerts2004,Chen2011}. The fundamental difficulty in achieving efficient coupling is the large modal mismatch in spatial extent, polarization, and propagation constant between a glass fiber with weak refractive index contrast and that of a typical on-chip waveguide with high refractive index contrast and sub-micron dimensions. A great variety of techniques have previously been developed to overcome this challenge, including end-fire coupling to inverse tapered on-chip waveguides~\cite{Mitomi1994,Almeida2003,McNab2003,Chen2010,Cohen2013}, on-chip gratings for near-normal incidence coupling~\cite{Taillaert2002,Masanovic2003,Chen2008}, and evanescent contra-directional coupling between tapered fibers and dispersion-engineered photonic crystal (PC) waveguides~\cite{Barclay2004}. End-fire coupling, while an efficient and mature technology, is limited to coupling at the edge of the chip and requires accurate three-dimensional alignment. Grating coupling and contra-directional PC waveguide coupling do allow for two-dimensional (2D) testing of arrays of devices on a chip, but suffer from a limited bandwidth of operation associated with phase-matching requirements.

In this letter we demonstrate optical fiber coupling to a 2D array of silicon photonic crystal cavities formed in the top device layer of a silicon-on-insulator (SOI) chip. The optical coupling in this work has the combined attributes of extremely high efficiency ($> 95\%$), large optical bandwidths ($> 50$~nm), and full in-plane device testing -- attributes which were critical to the recent demonstration of optical squeezing through radiation pressure effects in these mechanically compliant cavities~\cite{Safavi-Naeini2013b}. The coupling method involves using a tapered optical fiber in conjunction with a lithographically tapered, on-chip waveguide to achieve adiabatic mode transfer between fiber and chip. Since the coupling scheme utilizes neither phase- nor mode-matching, it achieves large bandwidths while being relatively robust to misalignment. This allows for rapid, repeatable, and wafer-scale optical fiber testing of devices. In addition to SOI, this method is applicable to photonic chips with a top thin-film (bottom cladding) of refractive index greater than (less than) the refractive index of the optical fiber.


\begin{figure*}[ht!]
\begin{center}
\includegraphics[width=1.6\columnwidth]{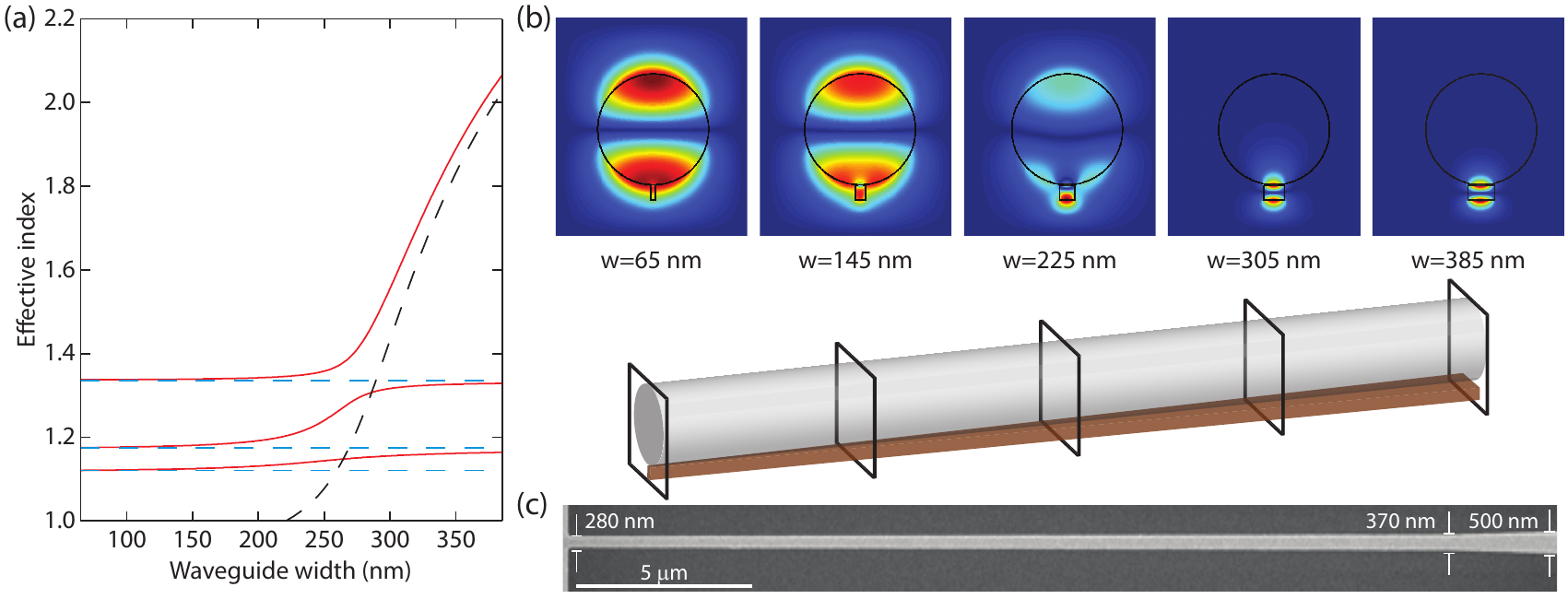}
\caption{\textbf{Fiber-to-waveguide adiabatic coupler.} (\textbf{a}) Simulation of the effective index of the three lowest-order modes of the bare optical fiber taper (dashed blue lines) and fundamental mode of the bare silicon width-tapered waveguide (dashed black line). The dispersion of the supermodes of the coupled waveguides are shown as solid red lines. The silicon waveguide is simulated with 220~nm thickness, while the optical fiber taper has a diameter of $1.6~\mu$m. (\textbf{b}) (top) Cross-sectional view of the supermode between fundamental fiber taper and fundamental silicon waveguide modes at various silicon waveguide widths. (bottom) Three-dimensional rendering of the fiber taper sitting on top of, and in contact with, the width-tapered silicon waveguide (sections corresponding to the simulated supermode mode profile are indicated). (\textbf{c}) Scanning electron micrograph (SEM) image of the waveguide made in the device layer of an SOI chip. The nominal width of the waveguide changes adiabatically from 280~nm to 370~nm over a distance of $25~\mu$m.}
\label{fig:1}
\end{center}
\end{figure*}

The concept of adiabatic transfer of energy between coupled modes dates back to the work of Landau and Zener~\cite{Zener1932} on molecular energy level crossings, ideas which were then applied to microwave~\cite{Cook1955,Fox1955,Louisell1955} and optical~\cite{Yajima1973,Burns1975,Silberberg1987,Dalgoutte1975,Shani1991} systems. A similar idea has recently been used to transfer light to a nanoscale cavity~\cite{Thompson2013}. In guided-wave optics, a direct analogy to the quantum mechanical formulation of adiabaticity can be found, where the effective index of the waveguide plays the role of energy levels in quantum mechanics, while the role of time is played by the propagation axis. In the adiabatic optical coupler of this work, a fiber taper with micron-scale diameter~\cite{Michael2007} is evanescently coupled to a silicon waveguide etched into the $220$~nm thick device layer of an SOI chip. These two waveguide systems, when brought into close proximity, lead to an effective interaction in the coupled mode picture. The effective refractive index of the on-chip silicon waveguide mode can be tailored in the interaction region by simply modifying the width of the waveguide along the propagation axis -- analogous to shifting energy levels in time in the quantum mechanics problem. By adiabatically sweeping the effective index of the silicon waveguide from below to above the effective index of the optical fiber ($n_f\lesssim 1.5$), all the energy originally in the silicon waveguide can be transferred to the optical fiber, and vice versa. 

This intuitive picture of the adiabatic coupling process is verified here through use of a vector finite-difference mode solver~\cite{Fallahkhair2008}. The mode solver is used to calculate the propagating modes of the coupled fiber-waveguide system for different silicon waveguide widths, which are plotted in Fig.~\ref{fig:1}a for an optical wavelength of $\lambda_0=1550$~nm. The uncoupled bands of the guided modes of a fiber taper of diameter $d=1.6$~$\mu$m and a silicon waveguide of thickness $t=220$~nm and varying width, $w$, are plotted as dashed lines in this plot. These uncoupled bands simply cross one another as the silicon waveguide width is swept from a large ($w\sim 385$~nm) to a small ($w\sim 65$~nm) width. For the most strongly coupled case, where the fiber and waveguide are touching, strong coupling between the bands is observed, forming mixed supermodes of the coupled waveguides. Following the highest effective refractive index supermode band as a function of waveguide width (or correspondingly, along the propagation axis), a drastic change in the supermode profile can be seen in Fig.~\ref{fig:1}b. For the smallest width waveguide, $w=65$~nm, the coupled mode is localized entirely in the optical fiber, whereas when $w=385$~nm the mode is entirely localized in the on-chip waveguide. 

In order for the coupling process to be adiabatic, and for near unity conversion efficiency of light from one waveguide to the other, the effective index of the fundamental mode of the silicon waveguide must be swept across the effective index of the fundamental optical fiber mode slowly. In exact analogy to the adiabatic theorem in quantum mechanics, the adiabatic condition can be expressed in terms of the size of the separation in the effective indices between the two closest bands, and the rate of change of the \textit{uncoupled} bands. The effective index of the fundamental silicon waveguide mode, $n_\text{WG}$, versus waveguide width is shown as a dashed black line in Fig.~\ref{fig:1}a. The relevant ``gap'' in the problem is $\Delta n_\text{eff}$, the effective index difference between the two supermodes formed from the fundamental fiber taper and fundamental silicon waveguide modes (see Fig.~\ref{fig:1}a). The adiabatic condition is then expressed as $\text{d} n_\text{WG}/\text{d}y \ll k_0 |\Delta n_\text{eff}|^2$, where $k_0$ is $2\pi$ divided by the free-space wavelength of light. For a fiber taper of slightly larger diameter $d\sim 2$~$\mu$m, as used in the experiments described below, the anti-crossing of the fundamental fiber taper and silicon waveguide modes shifts to a silicon waveguide width closer to $w\approx 325$~nm. A microfabricated waveguide meeting the adiabaticity condition and providing optimal coupling to a silica fiber taper of diameter $d=2$~$\mu$m is shown in Fig.~\ref{fig:1}c. In this structure the silicon waveguide width is varied from $w=280$~nm to $w=370$~nm over a distance of only $L=25~\mu$m. Finite-difference time-domain simulations of the coupling between the fundamental mode of the fiber taper and fundamental mode of the silicon waveguide show that the transfer efficiency of optical power from one waveguide mode to the other is $\gtrsim 97\%$ in the $1500$~nm wavelength band.


\begin{figure}[t]
\begin{center}
\includegraphics[width=.97\columnwidth]{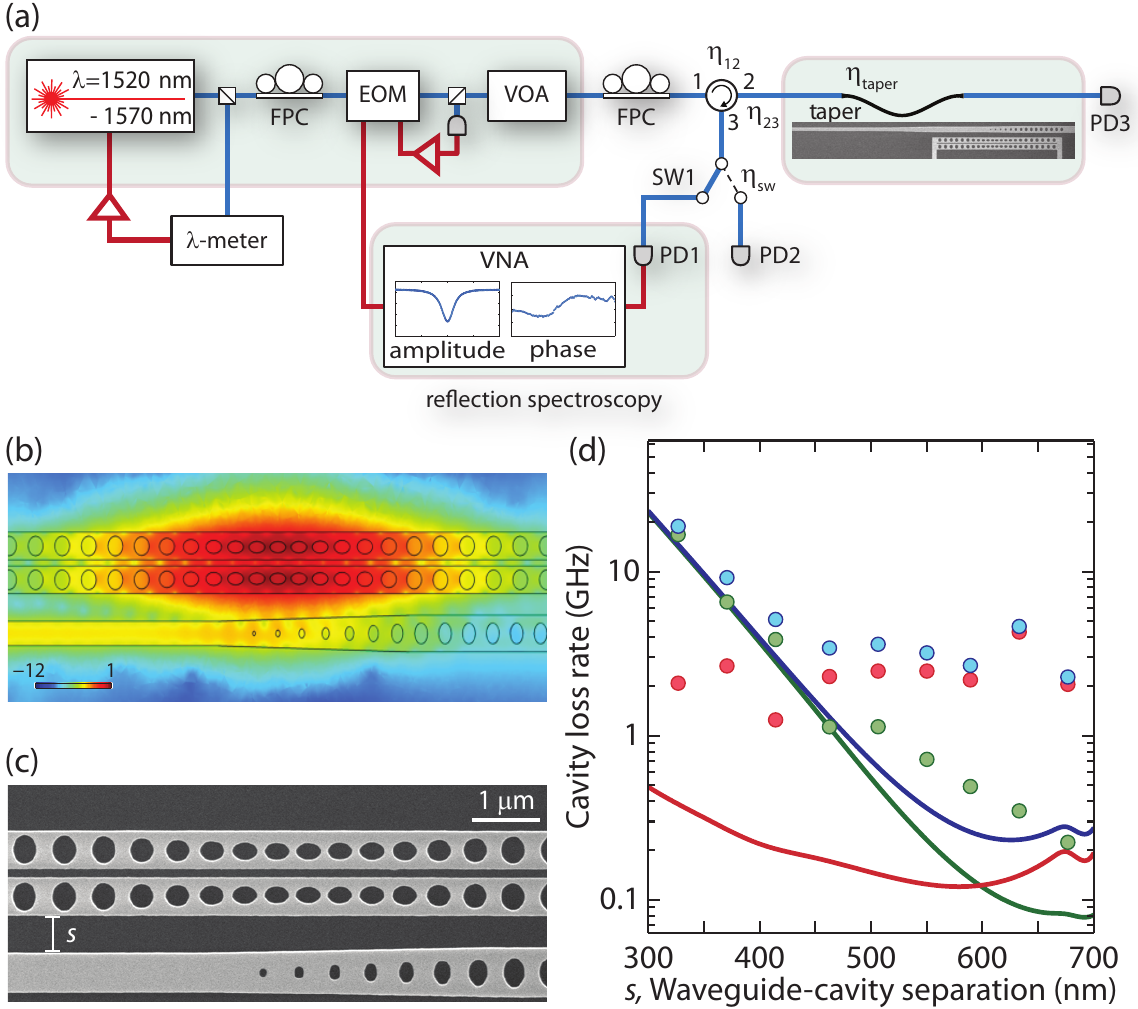}
\caption{\textbf{Experimental setup and waveguide-cavity loading.} (\textbf{a}) A tunable laser is used to characterize the devices and scan for optical resonances. The laser wavelength is stabilized at a fixed detuning from a cavity resonance using a wavemeter ($\lambda$-meter). The laser intensity is set using a variable optical attenuator (VOA) and amplitude modulated using an electro-optic modulator (EOM). The polarization of the laser light at the chip is set using a fiber polarization controller (FPC). A circulator is used to collect the reflected light from the chip. The reflected light passes through a switch (SW1), and can be either sent to a slow detector (PD2) to monitor the average reflected optical power, or to a high-speed detector (PD1). The high-speed detector is connected to a vector network analyzer (VNA) to analyze the phase and amplitude response of the optical cavity. The photodetector (PD3) is used to characterize the transmitted optical power if necessary. (\textbf{b}) Finite element methods (FEM) simulation of the optical cavity mode, field intensity normalized to its maximum, coupled to the waveguide. The waveguide is designed with a PC mirror which only allows light in and out of the cavity in one direction. (\textbf{c}) SEM of the waveguide-cavity coupling region of a fabricated device. (\textbf{d}) Simulation (solid lines) and experimentally extracted values (circles) for the intrinsic loss rate ($\ki/2\pi$; red), waveguide-cavity coupling rate ($\ke/2\pi$; green) and total cavity decay rate (($\kappa = \ki + \ke)/2\pi$; blue).}
\label{fig:2}
\end{center}
\end{figure}

Coupling of the on-chip silicon waveguide with the cavity is rather straightforward given the similarity of the material they are formed from. Waveguide-cavity coupling is accomplished by bringing the waveguide into the near-field from the side of the cavity, allowing for evanescent coupling. Here the cavity is a double nanobeam, zipper-style optomechanical cavity~\cite{Eichenfield2009a,Safavi-Naeini2013b}. A FEM model and SEM image of the waveguide-coupled cavity is shown in Fig.~\ref{fig:2}b and c, respectively. To ensure a single-sided coupling between the device and waveguide, the waveguide is terminated with a PC mirror~\cite{Chan2012}. The size of the holes of this mirror are adiabatically tapered to avoid scattering losses and the mirror region of the waveguide is the same as for the cavity itself. A plot of the theoretical cavity loss rates versus waveguide-cavity separation $s$ are plotted in Fig.~\ref{fig:2}d. In these plots the total cavity loss rate is $\kappa = \ki + \ke$, where $\ke$ is the waveguide-to-cavity coupling rate and $\ki$ is the rate associated with the internal losses of the cavity. $\ke$ (solid green line), is seen to decrease roughly exponentially with waveguide-cavity separation out to $s\approx 600$~nm, whereas $\ki$ (solid red line) is approximately independent of the gap.


Full characterization of the fiber-coupled cavity system can be separated into two parts:\ (i) characterization of waveguide-cavity loading and (ii) characterization of the fiber-to-waveguide coupling. The setup used for this characterization is shown in Fig.~\ref{fig:2}a. An external-cavity semiconductor laser (New Focus Velocity series) with a wavelength range of $1520-1570$~nm is used to characterize the cavity and coupler properties. An electro-optic modulator (EOM) is used to stabilize the optical power incident on the device and to create an optical sideband which allows for phase and amplitude heterodyne spectroscopy of the optical cavity (see below). Fiber polarization controllers (FPC) are used to set the polarization of the light incident on the device and a circulator allows for the collection of the reflected light. Further details are provided in the caption of Fig.~\ref{fig:2}.

\begin{figure}[t]
\begin{center}
\includegraphics[width=.86\columnwidth]{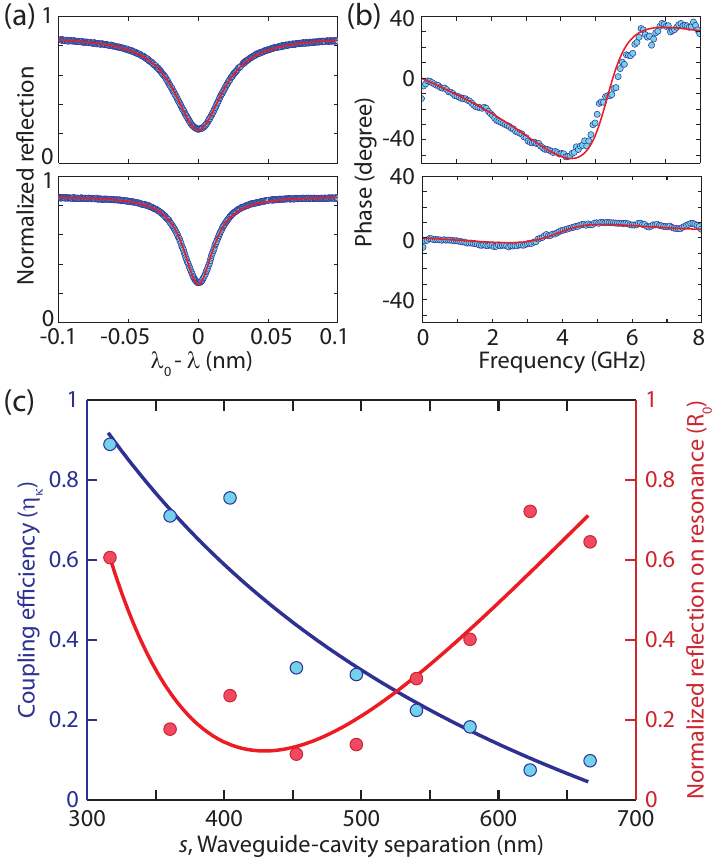}
\caption{\textbf{Waveguide-cavity coupling efficiency.} (\textbf{a}) Amplitude response of an over-coupled (top) and under-coupled (bottom) cavity. The over-coupled cavity has $R_0 = 0.26$ and $Q = 37,500$, while the under-coupled cavity has $R_0 = 0.30$ and $Q = 60,000$. \textbf{b} Phase response of the over-coupled (top) and under-coupled (bottom) cavities. (\textbf{c}) Normalized on-resonance reflection $R_0$ (red circles; right axis) and coupling efficiency $\eta_{\kappa}$ (blue circles; left axis) plotted as a function of waveguide-cavity separation $s$. Solid lines are a guide to the eye.}
\label{fig:3}
\end{center} 
\end{figure}

For the measurements presented here, an optical fiber taper formed from a Corning SMF28e fiber and drawn down to a diameter of $\approx 2~\mu$m~\cite{Knight1997} is positioned on top and in contact with the tapered region of the silicon waveguide using precision motorized stages. The fiber taper has a small radius ``dimple''~\cite{Michael2007} formed in it, providing a $10-15~\mu$m contact region with the tapered silicon waveguide. The efficiency of the fiber-waveguide coupling, $\eta_{c}$, can be determined by the off-resonance reflected optical power level referred to the known input power. Here the waveguide end mirror acts as a broadband reflector. Calibration of the circulator losses from port $1 \rightarrow 2$ ($\eta_{12}$) and port $2 \rightarrow 3$ ($\eta_{23}$), optical fiber taper loss ($\eta_{\mathrm{taper}}$), and the loss in optical switch SW1 ($\eta_{\text{SW}}$) allows the reflected power measured on the photodector PD2 to be used to determine $\eta_{c}$. This estimate of $\eta_{c}$ includes any scattering loss in the waveguide itself and non-perfect reflection from the waveguide end mirror. Figure~\ref{fig:3}a shows the reflected optical power as the laser is scanned over two different cavity resonances. The signal level is normalized to the input power such that off-resonance it corresponds to the efficiency of light coupled from the fiber taper into the waveguide and then back out again. In these scans, the background approaches a level of $\sim 0.85$ giving a single pass coupling efficiency of $\eta_c \sim 92$\% for both devices. The measured fiber-to-waveguide coupling varies slightly from device to device but is consistently over $90\%$ for single pass coupling, and in some cases $\eta_c$ has been measured larger than $95$\%. 

Characterization of the waveguide-cavity loading requires consideration of the amplitude and phase response of the cavity. The on-resonance reflection level, when normalized to the off-resonance level, is $R_0$ and is given by $R_0 = (1-2\eta_{\kappa})^2$ where $\eta_{\kappa}$ is the waveguide-cavity coupling efficiency and defined as $\eta_{\kappa} = \ke/\kappa$ for single-sided coupling. Fitting the linewidth of the optical response to determine $R_0$ and $\kappa$ is not enough to fully determine the coupling efficiency $\eta$, as $R_0$ is not a single-valued function of $\eta$ for single-sided coupling. This can be seen in Fig.~\ref{fig:3}a which shows the reflected signal from the optical scans of two different devices, both of which have similar $R_0$ values but different linewidths. The full, complex response of the optical cavity is measured by locking the laser off-resonance from the cavity, and using a vector network analyzer (VNA) in combination with an electro-optic modulator (see Fig.~\ref{fig:2}a) to sweep an optical sideband across the cavity. The beating of the probe sideband with the off-resonant laser signal is detected on a high-speed photodector (PD1) connected to the input of the VNA. By calibrating the response with no optical cavity present, the phase and amplitude response of the entire optical train is determined and used to normalize the signal. The amplitude and phase response of an over-coupled and under-coupled optical cavity measured using this technique are shown in Fig.~\ref{fig:3}a and b, respectively.

Figure~\ref{fig:3}c shows the inferred waveguide-cavity coupling efficiency for devices of varying gap size $s$. The right-hand axis shows the on-resonance reflection $R_0$ and the left-hand axis shows the equivalent waveguide-cavity coupling efficiency. Here the coupling efficiency increases roughly linearly with decreasing gap size and a maximum coupling efficiency approaching $\eta_{\kappa}=0.9$ is demonstrated. The measured $\ke$ and $\ki$ are also plotted as filled circles along with the theoretically expected values in Fig.~\ref{fig:2}d. The measured $\ke$ (filled green circles) follows a similar trend as the model, although with a slightly reduced exponential decay constant as a function $s$. The measured $\ki$ (filled red circles) is roughly an order of magnitude larger than the theoretical internal cavity losses, due primarily to surface roughness and absorption not captured in the simple model.


In summary, we have developed a method for coupling light from an optical fiber to an on-chip silicon waveguide with a demonstrated efficiency as high as $95\%$, and then used this on-chip waveguide to couple to a PC cavity with an efficiency of up to $90\%$. Taken together, this results in a total coupling efficiency of $85\%$ of photons emitted by the cavity and collected into the optical fiber. These sorts of fiber-to-cavity couplers are expected to find useful application for rapid wafer-scale testing of on-chip optical components~\cite{Chen2011}, and in quantum optics and optomechanics experiments with micro- and nano-optics devices~\cite{Srinivasan2007a,Safavi-Naeini2011a,Hausmann2012,Safavi-Naeini2013b} where highly efficient collection and detection of light is paramount.

\begin{acknowledgments}
The authors would like to thank J.\ D.\ Thompson and N.\ P.\ de Leon for valuable discussions. This work was supported by the AFOSR Hybrid Nanophotonics MURI, the DARPA/MTO ORCHID program through a grant from the AFOSR, and the Institute for Quantum Information and Matter, an NSF Physics Frontiers Center with support from the Gordon and Betty Moore Foundation. ASN and JC gratefully acknowledge support from NSERC. SG was supported by a Marie Curie International Outgoing Fellowship within the $7^{\textrm{th}}$ European Community Framework Programme.
\end{acknowledgments}

\def\urlprefix{}
\def\url#1{}

\end{document}